# Trade-off Energy and Spectral Efficiency in 5G Massive MIMO System

**Adeb Salh** [1*,] **Nor Shahida Mohd Shah** [2*], **Lukman Audah** [1], **Qazwan Abdullah** [1], **Norsaliza Abdullah** [3],
**Shipun A. Hamzah** [1,] **Abdu Saif** [1]

[1]Faculty of Electrical and Electronic Engineering, Universiti Tun Hussein Onn Malaysia, Parit Raja, Batu Pahat, Johor, Malaysia.
[2]Faculty of Engineering Technology, Universiti Tun Hussein Onn Malaysia, Pagoh, Muar, Johor, Malaysia.
[3]Faculty of Applied Sciences and Technology, Universiti Tun Hussein Onn Malaysia, Pagoh, Muar, Johor, Malaysia.
*Corresponding authors' e-mail: adeebsalh11@gmail.com, gazwan20060215@gmail.com

**ABSTRACT**

A massive multiple-input multiple-output (MIMO) system is very important to optimize the trade-off energy-efficiency (EE) and spectral-efficiency (SE) in fifth-generation cellular networks (5G). The challenges for the next generation depend on increasing the high data traffic in the wireless communication system for both EE and SE. In this paper, the trade-off EE and SE based on the first derivative of transmit antennas and transmit power in downlink massive MIMO system has been investigated. The trade-off EE-SE by using multi-objective optimization problem to decrease transmit power has been analyzed. The EE and SE based on constraint maximum transmits power allocation and a number of antennas by computing the first derivative of transmit power to maximize the trade-off EE – SE has been improved. From the simulation results, the optimum trade-off between EE and SE can be obtained based on the first derivative by selecting the optimal antennas with low cost of transmit power. Therefore, based on an optimal optimization problem is flexible to make trade-offs between EE-SE for distinct preferences

## 1. INTRODUCTION

Growing demand to achieve high data traffic, high-resolution video streams and smart communication in cellular networks depends on a promising candidate massive MIMO system. The massive MIMO technique is a key to improve the energy-efficiency (EE) and spectral-efficiency (SE) in the fifth-generation (5G) wireless communication networks. The deployment of cellular networks has a great potential to improve the low power base station (BS) and enhance the EE in a cellular network [1]-[6]. Due to the increasing attention in cellular networks, the performance of 5G depends on evaluating the EE in the massive MIMO system. A massive antenna array at the BS is able to provide the high connection to multiple single active users (UEs) in same time-frequency resource. The EE and SE cannot be simultaneously optimized. It is still quasi-concave due to noise amplifier, a large number of antennas, and cost of hardware, which decreases the radio frequency chains (RF) [7]-[15]. A large number of antennas in both linear precoder and decoder are needed to select the optimal antenna to limit the noise with uncorrelated interference. In addition, deploying more antennas causes higher circuit power consumption. From the perspective of energy-efficiency, the increasing base stations in the cellular networks reported that the total consumed energy in the whole network approximately 60-80% [16]-[20]. Adopted power scaling at transmitting signal from BS to users depends on reducing the interference. It takes into account the constraints of the transmit power allocation. Massive MIMO system helps to reduce transmit power created from high power gain and provides the higher EE. Moreover, distributed users in every cell and spectrum allocation are needed to improve spectral efficiency. From [21]-[26], the user association is optimized to enhance the EE by decreasing the transmit power consumption, while the area of spectral efficiency is linear based on a fixed number of users and an optimal number of antennas according to [27]. The degree of freedom in a massive MIMO system maximizes the SE by utilizing all available resources such as an available antenna that maximizes high data rate and transmit power. The area of SE growths linearly with a large number of transmit antennas, while with respect to user's association, the SE becomes a concave function. The future cellular networks must collect the explosive demand to increase the high data rate with low complexity of energy consumption.

From the existing works, the author in [28] studied the maximizing trade-off EE and SE based on user's association, number of antennas, power coordination and takes into account backhaul capacity to improve the performance EE-SE. Ensuring good rate fairness, lower-level power with user's association improves the performance of EE and SE in the [29]. Meanwhile, the author in [30] investigated the EE and SE based on K-tier heterogeneous networks by offloading data traffic to small cell massive MIMO system. The improving trade-off EE and SE with Rayleigh fading channel is based on using the different types of theoretical power consumption model and realistic power consumption model accordingly [31]. Another author investigated that the maximization of EE-SE in downlink massive MIMO system depends on analyzing the signal to interference noise ratio and optimum transmission power in each cell when the number of BSs for SE is high [32]. Achieving high sum-rates by using many transmit antennas, the balancing of EE-SE depends on admitting an optimal number of distributed users and active transmit antenna with low cost of power consumption [33]. The author in [34] focused on reducing the high complexity hybrid beamforming and power consumption for radio frequency chains by utilizing Butler phase shifter to enhance energy efficiency. From [35], the optimal design of EE-SE improved by employing the uplink and downlink high data rates under imperfect channel state information (CSI) combined with maximum ratio, matched filter and zero-forcing. Maximum EE-SE trade-off in massive MIMO system is achieved by equipping a great number of active users with less power and pilot training signal.

In this paper, accurate CSI depend on used time-division duplexing (TDD). Also, the goal of solving multi-objectivity problems is to find a set of trade-off solutions, by derivation of the EE-SE trade-off in closed form. To decrease the complexity based on computing, the essential derivative of EE and SE are in terms of power.

## 2. SYSTEM MODEL

Considering a downlink multiuser massive MIMO system, it is assumed that every cell contains one base station, every BS contains many transmit antenna $M$ and $K$ single users. It is assumed that $M \gg K$. The channel matrix between BS and users $\mathcal{H} = [h_1^T, h_2^T, \ldots . h_K^T]^T \in C^{1 \times M}$, as shown in Fig.1, with the assumption that the BS has perfect channel state information.

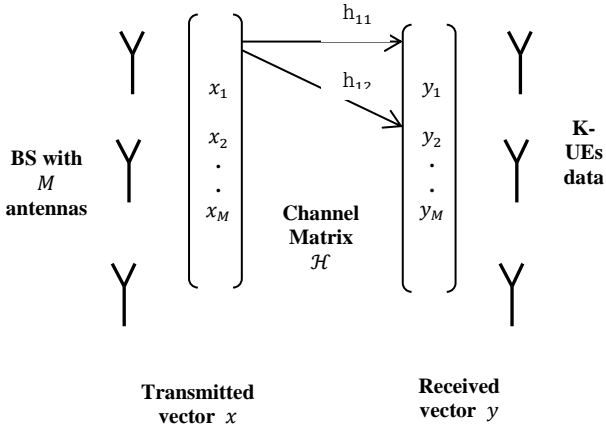

**Figure 1:** System model for multi-user massive MIMO system.

According to the increase in Shannon capacity, high data rate can be achieved by increasing the transmit power of channel to users. The zero-forcing beamforming precoding is used to mitigate the inter-user interference. The zero-forcing beamforming $V = [v_1, v_2, \ldots . v_K]$ can be written as

$$V = \mathcal{H}^H (\mathcal{H}\mathcal{H}^H)^{-1} \qquad (1)$$

The received signal $y_k$ of users $Kth$ inside $jth$ cell can be written as

$$y_k = \sqrt{\frac{\rho_d}{K\mathbb{F}}} h_k v_k x_k + \sqrt{\frac{\rho_d}{K\mathbb{F}}} \sum_{j=1,j\neq k}^{K} h_k v_j x_j + n_k$$

$$y_k = \sqrt{\frac{\rho_d}{K\mathbb{F}}} x_k + n_k \qquad (2)$$

where $h_k$ is the Hermitian transpose channel matrix, and $\rho_d$ is the transmit power in the downlink. The transmit signal vector $\phi_j \in C^{M \times 1}$ from BS $j$ a $M \times 1$ preceded vector $\phi_k = v_k x_k$, where each UE receives the signal vector from each BS, $v_k \in C^{M \times K}$ is the linear precoding matrix, $x_k \in C^K \sim \mathcal{CN}(0, I_K)$ is the data transmitted from BS to UEs. $\mathbb{F}$ is the normalization factor of $Kth$ UEs, which is expressed as $\mathbb{F} = \|V\|_f^2 / K$, $\|.\|_f^2$ represents the Frobenius norm and $n_k$ is the received noise with zero mean and variance.

### 2.1 Power Consumption

Reducing the transmit power depends on improving transmit antenna in a massive MIMO system. The circuit power consumption is established depend on selecting the optimal antennas at the BSs created due to antenna architectures considered not only from a power amplifier, but also from power consumption such as baseband processing, DAC, and filter. The total power consumption can be written as

$$Q_{max} = \rho_d + Q_C \qquad (3)$$

The circuit power consumption in cellular networks is continuously increasing based on a large distributed UEs and higher traffic demands. The circuit power consumption in [23]-[27],[36] is used and modeled as

$$Q_C \approx N(Q_{DAC} + Q_{mix} + Q_{filter}) + 2Q_{syn} + Q_{LNA} + Q_{IFA} + Q_{filter} + Q_{mix} + Q_{ADC} \qquad (4)$$

where $Q_{DAC}, Q_{mix}, Q_{filter}, Q_{syn}, Q_{LNA} + Q_{ADC}, Q_{filter}$ and $Q_{IFA}$ are the power consumption value for the DAC, the mixer, the filter in the transmitter, frequency synthesizer, the power consumption for soft noise amplifier, the power for the transitional frequency amplifier, and power consumed due to analog to digital converter, respectively. To simplify the power consumption, $Q_C = Q_1 + M_t Q_2$, where $Q_1 = 2Q_{syn} + Q_{LNA} + Q_{IFA} + Q_{filter} + Q_{mix} + Q_{ADC}$ and $Q_2 = Q_{DAC} + Q_{mix} + Q_{filter}$. Then, the total power consumption can be simplified as

$$Q_{max} = \rho_d + Q_C = \rho_d + Q_1 + NQ_2 \qquad (5)$$

### 2.2 Trade-off Energy Efficiency and Spectral Efficiency

In this section, the objectives are to optimize EE-SE of multi-user in terms of transmitting a huge number of antennas and transmit power. From (2), it is assumed that the capacity of $Kth$ UEs is multiplied by a bandwidth written as

$$C_k = \beta \log_2\left(1 + \frac{\rho_d}{\|V\|_f^2}\right) \quad (6)$$

where $\beta$ is the bandwidth. From (6), $\|V\|_f^2$ can be equivalent as $\|V\|_f^2 \approx \sum_{k=1}^{K} \frac{1}{\|h_k\|^2}$ at the number of antennas to serve a number of UEs $M \gg K$, where

$$\frac{1}{\|V\|_f^2} \approx \frac{1}{\sum_{k=1}^K \frac{1}{\|h_k\|^2}} \leq \frac{1}{K^2}\sum_{k=1}^K \|h_k\|^2 = \frac{1}{K^2}\sum_{j=1}^{M_t}\sum_{k=1}^K \|h_k\|^2 \quad (7)$$

The maximal SE is equivalent to limit the transmit power and a large number of transmit antennas for equivalent quality of the channel. From the Chi-square distribution for an increasing number of antennas $M_t$, where $\gamma_j = \sum_{k=1}^K \|h_{j,k}\|^2$, $j = 1, \ldots, M_t$ and $\gamma_1 \geq \gamma_2 \geq \ldots \geq \gamma_M$. There are large numbers of antennas that are equipped in a massive MIMO system which requires selecting the optimal antenna with low cost of transmit power. According to [24], the capacity of transmit antennas selection is expressed as

$$C_k = K\beta \log_2\left(1 + \frac{\rho_d}{K^2}\sum_{j=1}^N \gamma_N\right) \quad (8)$$

where $N$ is the selection of optimal antennas from available transmit antennas. From [25],[37], the mutual information for a channel with multiple receive antenna selection is

$$\mathbb{E}\left\{\sum_{j=1}^N \gamma_N\right\} = KN\left(1 + \ln\left(\frac{M_t}{N}\right)\right) \quad (9)$$

From (8) and (9), the average capacity can be written as

$$\mathbb{E}\{C_k\} = K\beta \log_2\left(1 + \frac{\rho_d N}{K}\right)\left(1 + \ln\left(\frac{M_t}{N}\right)\right) \quad (10)$$

A high achievable SE in per unit of bandwidth can be obtained by taking into account interference between users in a massive MIMO system. Distributing more than one user is scheduled in every cell to eliminate the inter-cell interference, and can be expressed as

$$\eta^{SE}{}_k = K \log_2\left(1 + \left(1 + \ln\frac{M_t}{N}\right)\frac{\rho_d N}{K}\right) \quad (11)$$

From (5) and (11) the maximal EE can be written as

$$\eta^{EE}{}_k = \frac{\eta^{SE}{}_k}{Q_{max}} = \frac{K\log_2\left(1+\left(1+\ln\frac{M_t}{N}\right)\frac{\rho_d N}{K}\right)}{\rho_d + Q_1 + NQ_2} \quad (12)$$

**2.3 Trade-off the EE-SE With Multi-Objective Optimization Problem**

The better EE-SE trade-off depends on achieving $\eta^{SE}{}_k$ for every UE and limits the transmit power at an increasing number of antennas $\eta^{SE}{}_k(\rho_d, N)$. From (11), the SE increases proportionally with transmitted power when the number of transmitting antennas $N$ is fixed [26]-[29],[38]. In addition, from (11) the performance of SE is obtained with a small and constant number of selected antennas $N$. The derivative of SE according to downlink transmit power $\rho_d$

$$\frac{\partial \eta^{SE}{}_k}{\partial \rho_d} = \frac{\partial\left(K\log_2\left(1+\left(1+\ln\frac{M_t}{N}\right)\frac{N\rho_d}{K}\right)\right)}{\partial \rho_d} = \frac{\left(1+\ln\frac{M_t}{N}\right)N}{\left(1+\left(1+\ln\frac{M_t}{N}\right)\frac{N\rho_d}{K}\right)\ln 2} \geq 0 \quad (13)$$

The performance SE in terms a number of transmit antennas can be expressed as

$$\frac{\partial \eta^{SE}{}_k}{\partial N} = \frac{\partial\left(K\log_2\left(1+\left(1+\ln\frac{M_t}{N}\right)\frac{N\rho_d}{K}\right)\right)}{\partial N} = \frac{\rho_d \ln\frac{M_t}{N}}{\ln 2\left(1+\left(\frac{N\rho_d}{K}\right)\left(1+\ln\frac{M_t}{N}\right)\right)} \geq 0 \quad (14)$$

From (14), the $\frac{\partial \eta^{SE}{}_k}{\partial N}$ strictly increases depending on selecting the optimal antenna from available antenna transmission, while the $\eta^{EE}{}_k$ firstly increases after decreasing with transmit power $\rho_d$ and fixed a number of transmit antennas $N$ in a massive MIMO system.

The performance constraints of the trade-off EE-SE for transmitting power may not be enough to maximize spectral efficiency or energy efficiency to obtain a comprehensive insight for both transmit antennas $N$ and power $\rho_d$ for EE [39]. In this case, the trade-off EE-SE by using multi-objective optimization problem to decrease transmit power is analyzed as follows:

$$\max_{Q_{max}, N}\left\{\eta^{SE}{}_k(\rho_d^{max}, N), \eta^{EE}{}_k(\rho_d^{max}, N)\right\} \quad (15)$$

S.t $K \leq N \leq \mathcal{N}$ (16)

$0 \leq \rho_d \leq \rho_d^{max}$ (17)

where $\mathcal{N}$ is the number of available transmit antennas and $\rho_d^{max}$ is the maximum transmit power. The trade-off EE-SE during transmit power $\rho_d = [0, \rho_d^{max}]$ the $\eta^{SE}{}_k(\rho_d) > \eta^{SE}{}_k(\rho_d^{max})$ and $\eta^{EE}{}_k(\rho_d) > \eta^{EE}{}_k(\rho_d^{max})$ the performance of $\eta^{EE}{}_k(\rho_d)$ increases when $\rho_d \geq \rho_d^{max}$ during $[0, \rho_d^{max}]$. A large number of transmit antennas provide more transmit power which causes a decreasing in EE according to (16). From (14) the $\eta^{SE}{}_k$ strictly increases with $N$ and $\rho_d$, while the $\eta^{EE}{}_k$ become concave shape when number of $N$ and $\rho_d$ increases. Also, from (17) the EE decreases when the transmitted power $\rho_d \leq \rho_d^{max}$ equals maximum transmit circuit power $Q_d^{max}$

$$\frac{\partial(\eta^{EE}{}_k)}{\partial \rho_d} = \frac{\partial\left(K\log_2\left(1+\left(1+\ln\frac{M_t}{N}\right)\frac{N\rho_d}{K}\right)\right)}{\rho_d + Q_1 + NQ_2} \quad (18)$$

We assume some of samples to simplify the following equations

$(\rho_d + Q_1 + NQ_2) = Q_{max}, \left(1 + \left(1 + \ln\frac{M_t}{N}\right)\frac{N\rho_d}{K}\right)\ln 2 = \Omega$

$$\left[\frac{\left(1+\ln\frac{M_t}{N}\right)N}{\Omega}\right](Q_{max}) - [K\log_2(\Omega)] \Big/ \left[((Q_{max},))^2\right] \quad (19)$$

$$\frac{\left[\left(1+\ln\frac{M_t}{N}\right)N\right](Q_{max},)-(\Omega)[K\log_2(\Omega)]}{(\Omega)\ln 2} \Big/ \left[((Q_{max},))^2\right] \quad (20)$$

$$\frac{\partial(\eta^{EE}_{\ k})}{\partial \rho_d} = \frac{N^2 \frac{\rho_d}{K}\left(\left(1+\ln\frac{M_t}{N}\right)\right)}{\left(1+\left(1+\ln\frac{M_t}{N}\right)\frac{N\rho_d}{K}\right)^2 \ln 2} < 0 \quad (21)$$

From the first derivative in (19), the EE increases and then decreases when the transmit power increases $\rho_d$. In addition, from the second term in the numerator $\lim_{\rho_d \to \infty}\left(\left(1+\left(1+\ln\frac{M_t}{N}\right)\frac{N\rho_d}{K}\right)\ln 2\right)\left[K\log_2\left(1+\left(1+\ln\frac{M_t}{N}\right)\frac{N\rho_d}{K}\right)\right] < 0$, while the first term $\lim_{\rho_d \to 0}\left[\left(1+\ln\frac{M_t}{N}\right)N\right]\left((\rho_d+Q_1+NQ_2)\right) > 0$, the EE starts increasing and after that starts decreasing due to large cost for every antenna that contains the radio frequency chains and power amplifier. From (21) maximizing EE depends on obtaining the optimal antenna selection and optimal transmit power allocation.

In case of $\eta^{SE}_{\ k} = \min \eta^{SE}_{\ k}$: from (15) the relation of EE strict decreases with SE, while the EE become maximized when the SE is small $\eta^{SE}_{\ k} = \min \eta^{SE}_{\ k}$

$$\left.\frac{\partial \eta^{EE}_{\ k}(\eta^{SE}_{\ k})}{\partial \eta^{SE}_{\ k}}\right|_{\eta^{SE}_{\ k}=\eta^{SE}_{\ k}(\min)} \leq 0 \quad (22)$$

From (22) the EE firstly increases with minimum SE based on available antenna selection and transmits power.

In another case, when $\eta^{SE}_{\ k} = \eta^{SE}_{\ k}(\max)$, the EE $\eta^{EE}_{\ k}$ firstly increases with SE and can be a performance to be maximized when the power allocation become allocated to every UE at a fixed number of transmit antennas as follows;

$$\left.\frac{\partial \eta^{EE}_{\ k}(\eta^{SE}_{\ k})}{\partial \eta^{SE}_{\ k}}\right|_{\eta^{SE}_{\ k}=\eta^{SE}_{\ k}(max)} \geq 0 \quad (23)$$

### 3. NUMERICAL RESULTS

In this section, the numerical results are presented by using MATLAB based on the Monte Carlo simulations to verify the theoretical analysis.

From Fig. 2, there is a trade-off between EE and the number of antennas $N$, the use of different circuits power, the hardware cost of circuit power created from noise amplifier, DAC and radio frequency chains that impact of $Q_C$ that limit the EE. The maximum value of EE is obtained by distributing a number of users and fixing the transmit power. The EE starts increasing when the number of available antennas is small, while a large number of transmit antennas makes the EE start decreasing due to the large number of RF chains and consumption of circuit power according to $Q_{max} = \rho_d + Q_1 + NQ_2$ in (5). The $\eta^{EE}_{\ k}$ becomes concave shape when number of $N$ and $\rho_d$ increases. From Fig. 2, the maximum EE = 32 Mbit/J, when the number of evaluable antennas $N = 20$, at fixed transmit power $\rho_d = 30$ dBm and number of distributed users $K = 8$, while increasing the number of served users $K = 15$, and fixed number of transmit power $\rho_d = 30$ dBm, the EE becomes small and equal to EE = 26 Mbit/J when the number of antennas $N = 18$.

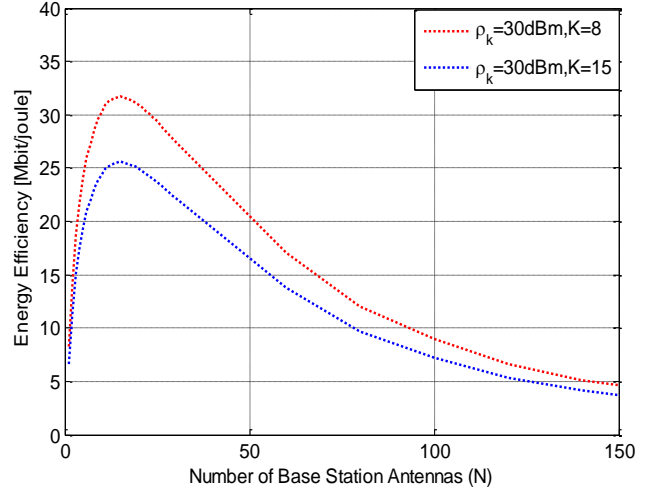

**Figure 2:** EE versus number of transmit antennas $N$

Figure 3 presents the trade-off between EE and SE. The transmit power and an available number of antennas are able to maximize the trade-off EE-SE by setting multi-objective optimization problem.

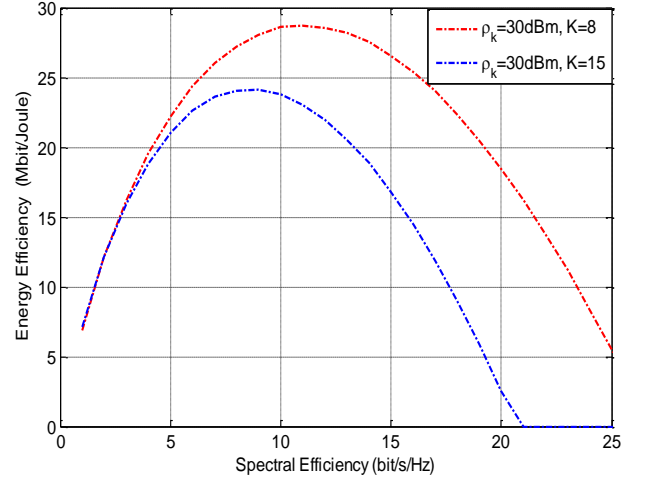

**Figure 3:** Trade of EE-SE in terms of power allocation and distributed a number of users.

The achievements of optimizing values for EE and SE is based on the number of distributed users with fixed transmit power. The trade-off between EE and SE increases simultaneously with a first corresponding minimum SE to the number of employed antennas with less transmit power. The $\eta^{EE}_{\ k}$ becomes quasi-concave in the region $\left[\eta^{SE}_{\ k}(\min), \eta^{SE}_{\ k}(\max)\right]$ according to (22) and (23). The performance of $\eta^{EE}_{\ k}(\rho_d)$ increases when $\rho_d \leq \rho_d^{\max}$. Selecting the optimal transmit power $\rho_d = 30$ dBm, and a number of the employed active users K = 8, the EE increases with the SE and gives the value EE = 28 Mbit/J, at SE = 12 bit/s/Hz. Meanwhile, when the optimal transmit power $\rho_d$ = 30 dBm, at a distributed number of active users increases to K = 15, the EE = 23 Mbit/J with SE = 8bit/s/Hz and become small values due to a large transmit power. After this value, the EE starts to decrease with SE due to the complex processing complexity and high operation cost for radio frequency chains at the employed antenna in a massive MIMO system.

## 4. CONCLUSION

In this paper, the optimization of EE–SE trade-off in massive MIMO system with first derivative into multi-objectivity problems for number of available antenna and transmit power in downlink massive MIMO system has been investigated. The trade-off EE - SE is based on a number of transmit antennas and transmit power by using multi-objective optimization problem to decrease transmit power. Simulation results indicate that the optimization of the EE and SE is achieved based on minimizing the total energy consumption with proportional transmitting power allocation when the number of transmitted antennas *N* is fixed.

## ACKNOWLEDGEMENT

This research is supported by **Universiti Tun Hussein Onn Malaysia under the TIER1 Grant Vot H243**.